\documentclass[aps,eqsecnum,showpacs,amsmath,amssymb]{revtex4}
\usepackage{graphics}

\begin{document}

\newcommand{\xbf}[1]{\mbox{\boldmath $ #1 $}}

\newcommand{\sixj}[6]{\mbox{$\left\{ \begin{array}{ccc} {#1} & {#2} &
{#3} \\ {#4} & {#5} & {#6} \end{array} \right\}$}}

\newcommand{\threej}[6]{\mbox{$\left( \begin{array}{ccc} {#1} & {#2} &
{#3} \\ {#4} & {#5} & {#6} \end{array} \right)$}}

\newcommand{\clebsch}[6]{\mbox{$\left( \begin{array}{cc|c} {#1} & {#2} &
{#3} \\ {#4} & {#5} & {#6} \end{array} \right)$}}

\newcommand{\iso}[6]{\mbox{$\left( \begin{array}{cc||c} {#1} & {#2} &
{#3} \\ {#4} & {#5} & {#6} \end{array} \right)$}}

\preprint{{DOE/ER/40762-305}\cr{UMPP\#04-031}}

\begin{flushright}

DOE/ER/40762-325\\
UMPP\#05-016
\end{flushright}

\count255=\time\divide\count255 by 60
\xdef\hourmin{\number\count255}
  \multiply\count255 by-60\advance\count255 by\time
 \xdef\hourmin{\hourmin:\ifnum\count255<10 0\fi\the\count255}

\title{SU(3) Clebsch-Gordan Coefficients for Baryon-Meson Coupling at
Arbitrary $N_c$}


\author{Thomas D. Cohen}
\email{cohen@physics.umd.edu}



\affiliation{Department of Physics, University of Maryland,
College Park, MD 20742-4111}

\author{Richard F. Lebed}
\email{Richard.Lebed@asu.edu}


\affiliation{Department of Physics and Astronomy, Arizona State
University, Tempe, AZ 85287-1504}

\date{October, 2004}

\begin{abstract}
We present explicit formul{\ae} for the SU(3) Clebsch-Gordan
coefficients that are relevant for the couplings of large $N_c$
baryons to mesons.  In particular, we compute the Clebsch-Gordan
series for the coupling of the octet (associated with mesons, and
remains the correct representation at large $N_c$) to the large $N_c$
analogs of the baryon octet and decuplet representations.
\end{abstract}


\maketitle

\section{Introduction \label{sec:intro}}

The large $N_c$ limit of QCD and the $1/N_c$ expansion have
proven to be valuable methods to obtain qualitative insights into
the structure and interactions of mesons~\cite{Hoo} and
baryons~\cite{Wit}.  The approach became quantitatively valuable
in the study of baryon properties when it was realized that large
$N_c$ consistency conditions impose on the low-lying baryonic
states a contracted SU($2N_f$) spin-flavor symmetry that holds up
to corrections subleading in $1/N_c$~\cite{GS,DM}. This symmetry
allows one to relate certain baryonic properties in a
model-independent way, thereby providing real predictive power.

The best known applications of this method have occurred in the
study of properties of the ground-state band of baryons---those
baryons (such as the $\Delta$) that become degenerate with the
nucleon in the large $N_c$ limit of QCD.  These states all fall
into a single representation of contracted SU(2$N_f$), providing
(for example) a field-theoretical explanation for
group-theoretical results of the old SU(6) spin-flavor symmetry.
This means that matrix elements of {\em any\/} operator
contributing at leading order in the $1/N_c$ expansion between
{\em any\/} states in the band can be related to matrix elements
of the same operator between other states in the band, by purely
group-theoretical means.  Corrections are typically $O(1/N_c)$,
but for certain cases they are only $O(1/N_c^2)~$\cite{DM}.  This
allows for concrete, albeit approximate, predictions, which
appear to work at the level one would estimate from the size of
the neglected corrections~\cite{JL}.  This general formalism has
been developed both for two and three flavors~\cite{DJM1,DJM2}.
Of course, when working with three flavors, one must include the
effects of SU(3) flavor breaking, which can yield corrections
comparable to $1/N_c$ corrections~\cite{JL}.

At a technical level, the symmetry has been implemented in a number of
ways.  One is to solve algebraically the commutation relations that
arise from the consistency conditions~\cite{GS,DM}. A particularly
elegant method for accomplishing this is via the formalism of induced
representations~\cite{DJM1}.  An alternative, and somewhat more
pragmatic approach has been to map the full commutation relations onto
those of a simple constituent quark model, and then to solve using
quark model operators.  Since it is straightforward to do the $N_c$
counting for all of the operators that arise in the quark model, one
can quickly and efficiently derive the results~\cite{DJM2} of the
symmetry. Although this approach is expressed in quark model language,
it does not rely on any dynamical assumptions of the quark model.

Recently, a number of groups have developed techniques to extend
this analysis to excited states of the
baryons~\cite{CGKM,PY,Goity,CCGL1,CCGL2,CC,CL,PS,GSS}. This is an
important problem. Although the properties of excited baryons
have been studied in models for a very long time, the connection
of the models to QCD remains obscure.  Thus, there is a need for
reliable model-independent predictions tied directly to QCD;
large $N_c$ analysis provides such a tool.  The methods used so
far include a direct implementation of the consistency rules on
excited states~\cite{PY}, an operator analysis in the quark model
language~\cite{CGKM,Goity,CCGL1,CCGL2,CC,PS,GSS}, and a new
approach based on the study of resonances in the context of
model-independent relations between meson-baryon scattering
amplitudes~\cite{CL}.  The last approach has a important advantage
over the others, in that it does not implicitly assume excited
states to be stable at large $N_c$. This is significant since
widths of excited baryons are generically $O(N_c^0)$, and there
is no reason to assume that this generic behavior does not apply
to the excited states of interest~\cite{CDLN}.

To date, this general analysis has been confined to studies of two
flavors, although recently large $N_c$ analysis has been performed for
some exotic states of fixed strangeness~\cite{CLpenta} and for exotic
baryons containing heavy quarks~\cite{exotic}.  Of course, it is not
completely surprising that the two-flavor case was explored before the
three-flavor case, since generically it is easier to work with flavor
SU(2) than flavor SU(3). Indeed, this feature is particularly
prominent for studies at large $N_c$ QCD.  The reason for this is
rather simple: The SU(2) flavor representations that arise for
low-lying states of baryons in large $N_c$ QCD are identical to the
representations that occur for $N_c \! = \! 3$, while the SU(3) flavor
representations that arise for low-lying states of baryons in large
$N_c$ QCD are quite different from the representations that occur for
$N_c \! = \!  3$~\cite{DJM1,DP}.  This in turn means that explicit
dependence on $N_c$ can arise in matrix elements directly from the
$N_c$ dependence of the representation and not just from the operator
in question.  As noted in Ref.~\cite{tdc} in the context of exotic
states, the explicit $N_c$ dependence in the matrix elements can be
critical in obtaining the correct $N_c$ scaling of an observable.

To make progress on the general three-flavor problem using either the
direct large $N_c$ consistency condition method of Ref.~\cite{PY} or
via the model-independent meson-baryon scattering amplitudes of
Ref.~\cite{CL}, one needs to possess the relevant SU(3) Clebsch-Gordan
coefficients for the coupling of a flavor octet meson with the
representation of the baryon of interest.  Such coefficients are also
of use in computing matrix elements in the quark model basis. While
SU(3) Clebsch-Gordan tables exist for fairly large
representations~\cite{Kaeding1} and computer codes exist for
generating them for arbitrary (but fixed)
representations~\cite{Kaeding2}, it is extremely useful in conducting
large $N_c$ analysis to have explicit formul{\ae} for the relevant
coefficients in a form that makes the $N_c$ dependence manifest.  The
purpose of the present paper is to present such formul{\ae} to
facilitate progress in the field.

As noted above, that since baryons for arbitrary $N_c$ contain $N_c$
valence quarks, the corresponding baryon SU(3) representations also
grow in size with $N_c$~\cite{DJM1,DP}.  We wish to identify these
large $N_c$ representations with their $N_c \! = \! 3$ counterparts.
To keep our notation simple and aid in the extrapolation to $N_c \! =
\!  3$, we use quotes to denote the generalized SU(3) representations
familiar from $N_c \!  = \! 3$:
\begin{eqnarray}
\mbox{\rm ``1''} & \equiv & [ \, 0, (N_c \! - \! 3)/2 \, ] \ ,
\nonumber \\
\mbox{\rm ``8''} & \equiv & [ \, 1, (N_c \! - \! 1)/2 \, ] \ ,
\nonumber \\
\mbox{\rm ``10''} & \equiv & [ \, 3, (N_c \! - \! 3)/2 \, ] \ ,
\nonumber \\
``\overline{10}\mbox{''} & \equiv & [ \, 0, (N_c \! + \! 3)/2 \,
] \ ,
\nonumber \\
\mbox{\rm ``27''} & \equiv & [ \, 2, (N_c \! + \! 1)/2 \, ] \ ,
\nonumber \\
\mbox{\rm ``35''} & \equiv & [\, 4, (N_c \! - \! 1)/2 \, ] \ ,
\end{eqnarray}
while representations with that do not appear for $N_c \! = \! 3$
are left in the standard $(p,q)$ weight form.

In this work we focus on those representations that can decay into the
ordinary ground-state band baryons by the emission of an octet meson.
Since in the real world of $N_c \! = \! 3$, the only ground-state
baryons that occur are the ``8'' and ``10'', we present only the
Clebsch-Gordan coefficients for those representations that couple to
the ``8'' and ``10'' via an 8.



\section{Conventions and Method of Calculation}

Following the usual convention, we write the SU(3) Clebsch-Gordan
coefficients (CGCs) as products of ordinary SU(2) CGCs and {\it
isoscalar factors}. Isoscalar factors~\cite{EdmondsI} are the portions
of SU(3) CGCs that do not depend upon the isospin additive quantum
numbers $I_z$. The full SU(3) CGC is factored into a product of the
pure isospin SU(2) CGC and an isoscalar factor:
\begin{equation}
\clebsch{R_1}{R_2}{R_\gamma}{I_1,I_{1z},Y_1}{I_2,I_{2z},Y_2}{I,I_z,Y}
= \iso{R_1}{R_2}{R_\gamma}{I_1,Y_1}{I_2,Y_2}{I,Y}
\clebsch{I_1}{I_2}{I}{I_{1z}}{I_{2z}}{I_z} \ ,
\end{equation}
where the label $R$ indicates the SU(3) representation, which may be
denoted using the usual weight diagram notation $(p,q)$.  $\gamma$
labels degenerate representations occurring in a given product.  Since
both the full SU(3) and SU(2) CGCs form unitary matrices, so do
the isoscalar factors.

Presented here are tables of isoscalar factors for the products
\begin{eqnarray}
\mbox{\rm ``8''} \otimes 8 & = & \mbox{\rm ``27''} \oplus \mbox{\rm
``10''} \oplus ``\overline{10}\mbox{\rm ''} \oplus ``8_1\mbox{\rm ''}
\oplus ``8_2\mbox{\rm ''} \oplus ``1\mbox{\rm ''} \oplus
[ \, 2, (N_c \! - \! 5)/2 \, ] \ , \nonumber \\
\mbox{\rm ``10''} \otimes 8 & = & \mbox{\rm ``35''} \oplus
\mbox{\rm ``27''} \oplus ``10_1\mbox{\rm ''}
\oplus ``10_2\mbox{\rm ''} \oplus ``8\mbox{\rm ''} \oplus
[ \, 5,(N_c \! - \! 5)/2 \, ] \oplus [ \, 2,(N_c \! - \! 5)/2 \, ]
\oplus [ \, 4,(N_c \! - \! 7)/2 \, ] \ . \label{prods}
\end{eqnarray}
Factors for product representations that do not occur for $N_c \! = \!
3$ are suppressed because they are not useful in the phenomenological
analysis.  As a result, in some cases the matrices presented below
represent only a corner of a fully unitary matrix.

The SU(3) CGCs are computed by essentially the same procedure as SU(2)
CGCs: One uses the matrix elements of the raising and lowering
operators.  In the case of SU(3), these are given by~\cite{Bied,Hecht}
\begin{eqnarray}
U_+ | \, (p,q) \, I I_z Y \rangle & = & +g[(p,q), \, I, -I_z, \, Y] \
| \, (p,q), \, I \! + \! 1/2, \, I_z \! - \! 1/2, \, Y \! \! + \! 1 \,
\rangle \nonumber \\ & &
-g[(p,q), \, -(I \! + \! 1), \, -I_z, \, Y] \ | \, (p,q), \, I \! - \!
1/2, \, I_z \! - \! 1/2, \, Y \! \! + \! 1 \, \rangle \ , \nonumber \\
U_- | \, (p,q) \, I I_z Y \rangle & = & -g[(p,q), \, -(I \! + \! 3/2),
-(I_z \! + \! 1/2), \, Y \! - \! 1] \ | \, (p,q),
\, I \! + \! 1/2, \, I_z \! + \! 1/2, \, Y \! - \! 1 \, \rangle
\nonumber \\ & &
+g[(p,q), \, I \! - \! 1/2, \, -(I_z \! + \! 1/2), Y \! - \! 1] \ |
\, (p,q), \, I \! - \! 1/2, \, I_z \! + \! 1/2, \, Y \! - \! 1 \,
\rangle \ , \nonumber \\
V_+ | \, (p,q) \, I I_z Y \rangle & = & +g[(p,q), \, I, I_z, Y] \ |
\, (p,q), \, I \! + \! 1/2, \, I_z \! + \! 1/2, \, Y \! \! + \! 1 \,
\rangle \nonumber \\ & &
+g[(p,q), \, -(I \! + \! 1), \, I_z, \, Y] \ | \, (p,q), \, I \! - \!
1/2, \, I_z \! + \! 1/2, \, Y \! \! + \! 1 \, \rangle \ , \nonumber \\
V_- | \, (p,q) \, I I_z Y \rangle & = & +g[(p,q), \, -(I \! + \! 3/2),
I_z \! - \! 1/2, \, Y \! - \! 1] \ | \, (p,q),
\, I \! + \! 1/2, \, I_z \! - \! 1/2, \, Y \! - \! 1 \, \rangle
\nonumber \\ & &
+g[(p,q), \, I \! - \! 1/2, \, I_z \! - \! 1/2, Y \! - \! 1] \ |
\, (p,q), \, I \! - \! 1/2, \, I_z \! - \! 1/2, \, Y \! - \! 1 \,
\rangle \ ,
\end{eqnarray}
where the function $g$ is given by
\begin{equation}
g[(p,q) \, I I_z Y] = \left\{ \frac{(I \! + \! I_z \! + \! 1)[\frac 1
3 (p \! - \! q) \! + \! I \! + \! \frac Y 2 \! + \! 1][\frac 1 3 (p \!
+ \! 2q) \! + \! I \! + \! \frac Y 2 \! + \! 2][\frac 1 3 (2p \! + \!
q) \! - \! I \! - \! \frac Y 2]}{(2I \! + \! 1)(2I \! + \! 2)}
\right\}^{1/2} \ .
\end{equation}
One obtains recursion relations among SU(3) CGCs by combining two
states (the usual Clebsch-Gordan series) and computing matrix elements
of $U_{\pm} \! = \!  U_{1 \pm} \! + \! U_{2 \pm}$ and $V_{\pm} \! = \!
V_{1 \pm} \! + \!  V_{2 \pm}$.  Of course, this is entirely analogous
to the manner in which one computes SU(2) CGCs, in that case using
$T_{\pm}$ matrix elements.  The recursion relations in the SU(3) case
are however much more involved, generically involving 6 CGCs rather
than the 3 of the SU(2) case.  For example, the recursion relation for
the generator $V_+$ reads
\begin{eqnarray}
\lefteqn{\langle (p_1, q_1) I_1 I_{1z} Y_1 |
\langle (p_2, q_2) I_2 I_{2z} Y_2 |
V_+ | (p, q) I I_z Y \rangle} & & \nonumber \\
& = & +g[(p,q), I, I_z, Y]
\iso{(p_1,q_1)}{(p_2,q_2)}{(p,q)}{I_1 Y_1}{I_2 Y_2}
{I \! + \! \frac 1 2, Y \! + \! 1}
\clebsch{I_1}{I_2}{I \! + \! \frac 1 2}{I_{1z}}{I_{2z}}
{I_z \! + \! \frac 1 2} \nonumber \\
& & +g[(p,q), -(I \! + \! 1), I_z, Y]
\iso{(p_1,q_1)}{(p_2,q_2)}{(p,q)}{I_1 Y_1}{I_2 Y_2}
{I \! - \! \frac 1 2, Y \! + \! 1}
\clebsch{I_1}{I_2}{I \! - \! \frac 1 2}{I_{1z}}{I_{2z}}
{I_z \! + \! \frac 1 2} \nonumber \\
& = & +g[(p_1,q_1), -(I_1 \! + \! 3/2), \, I_{1z} \! - \! 1/2, \,
Y_1 \!  - \! 1]
\iso{(p_1,q_1)}{(p_2,q_2)}{(p,q)}{I_1 \! + \! \frac 1 2, \,
Y_1 \! - \! 1}{I_2 Y_2}{I Y}
\clebsch{I_1 \! + \! \frac 1 2}{I_2}{I}{I_{1z} \! - \! \frac 1 2}
{I_{2z}}{I_z} \nonumber \\
& & +g[(p_1,q_1), \, I_1 \! - \! 1/2, \, I_{1z} \! - \! 1/2, \, Y_1 \!
- \! 1]
\iso{(p_1,q_1)}{(p_2,q_2)}{(p,q)}{I_1 \! - \! \frac 1 2, \, Y_1 \! -
\! 1}{I_2 Y_2}{I Y}
\clebsch{I_1 \! - \! \frac 1 2}{I_2}{I}{I_{1z} \! - \! \frac 1 2}
{I_{2z}}{I_z} \nonumber \\
& & +g[(p_2,q_2), -(I_2 \! + \! 3/2), \, I_{2z} \! - \! 1/2, \,
Y_2 \!  - \! 1]
\iso{(p_1,q_1)}{(p_2,q_2)}{(p,q)}{I_1 Y_1}{I_2 \! + \! \frac 1 2, \,
Y_2 \! - \! 1}{I Y}
\clebsch{I_1}{I_2 \! + \! \frac 1 2}{I}{I_{1z}}
{I_{2z} \! - \! \frac 1 2}{I_z} \nonumber \\
& & +g[(p_2,q_2), \, I_2 \! - \! 1/2, \, I_{2z} \! - \! 1/2, \, Y_2 \!
- \! 1]
\iso{(p_1,q_1)}{(p_2,q_2)}{(p,q)}{I_1 Y_1}
{I_2 \! - \! \frac 1 2, \, Y_2 \! - \! 1}{I Y}
\clebsch{I_1}{I_2 \! - \! \frac 1 2}{I}{I_{1z}}
{I_{2z} \! - \! \frac 1 2}{I_z} \ .
\end{eqnarray}
Therefore, one must be much more judicious in the choice of quantum
numbers than in the SU(2) case in order to obtain relations that
contain a minimum number of unknown CGCs.  In practice, one begins in
the small ``corners'' of the SU(3) weight diagrams, where fewer states
contribute to the recursion relations; this is analogous to starting
with ``stretched'' states in the SU(2) case.

Since the recursion relations are homogeneous in CGCs, the next
necessary step is to obtain normalization conditions.  Here one uses
the unitarity of the CGC matrices, not only to set the scale of the
CGCs, but also to obtain some of the coefficients via orthogonality
conditions.  Finally, in order to obtain a unique value for the CGCs,
one must impose a phase condition.  We use the standard
Condon-Shortley~\cite{CS} convention for the SU(2) CGCs, which amounts
to
\begin{equation}
\clebsch{I_1}{I_2}{I}{I_1}{I \! - \! I_1}{I} > 0 \ .
\end{equation}
The phase of the SU(3) CGCs is determined by the de~Swart
convention~\cite{deSwart}, which amounts to
\begin{equation} \label{deSwcond}
\iso{R_1}{R_2}{R_\gamma}{\tilde{I}_1 \tilde{Y}_1}{\tilde{I}_2, Y \!
- \! \tilde{Y}_1}{I_{\rm max} \, Y} > 0 \ .
\end{equation}
Here, $I_{\rm max}$ is the highest value of isospin in the
multiplet $R_\gamma$ (and its selection uniquely fixes $Y$),
$\tilde{I}_1$ is the largest value of isospin in $R_1$ that
couples to this highest-weight state (and has corresponding
hypercharge $\tilde{Y}_1$), and $\tilde{I}_2$ is the largest
value of isospin in $R_2$ that couples $(\tilde{I}_1,
\tilde{Y}_1)$ to $(I_{\rm max} \, Y)$.  However, even this level
of description does not completely specify all phases in the case of
degenerate product representations (multiple allowed values of
$\gamma$).  In that case, the convention used
here~\cite{Kaeding1,Kaeding2} is defined as follows: If there are
$\Gamma$ values of $\gamma$, then $R_{\gamma = 1}$ is defined as the
representation for which
\begin{equation} \label{Kaedcond}
\iso{R_1}{R_2}{R_\gamma}{\tilde{I}_1 \tilde{Y}_1}{I_2, Y \!
- \! \tilde{Y}_1}{I_{\rm max} \, Y} = 0
\end{equation}
for the $\Gamma \! - \! 1$ highest allowed values of $I_2$, namely,
$\tilde{I}_2, \tilde{I}_2, \ldots, \tilde{I}_2 \! - \! \Gamma \! + \!
2$, but is positive for $I_2 \! = \! \tilde{I}_2 \! - \! \Gamma \!  +
\! 1$; this is sufficient to determine uniquely the isoscalar factors
for $R_{\gamma = 1}$.  Next, $R_{\gamma = 2}$ is defined to be the
representation orthogonal to $R_{\gamma = 1}$ such that
Eq.~(\ref{Kaedcond}) holds for the highest $\Gamma \!  - \! 2$ allowed
values of $I_2$ but is positive for $I_2 \! = \! \tilde{I}_2 \! - \!
\Gamma \!  + \! 2$, and so on.  Only for $R_{\gamma = \Gamma}$ in the
degenerate case does Eq.~(\ref{deSwcond}) hold at face value.

Note that $10_1$ defined in this way vanishes for $N_c \! = \! 3$.
Therefore, the tables contain only the representation $10_2$.  The
tables are designed to resemble as closely as possible those of
de~Swart~\cite{deSwart}; hence, we use there the notation
$\mu_\gamma$ instead of $R_\gamma$.  Indeed, they have been
checked against tables in this reference for $N_c \! = \! 3$, and
against the results from a computer program~\cite{Kaeding2} for
$N_c \! = \! 5$.  Finally, in the standard convention for
distinguishing degenerate representations, the current products
induce a couple of more complicated factors, which we abbreviate:
\begin{eqnarray}
       D  & \equiv & 5N_c^2+22N_c+9 \ , \\
\tilde{D} & \equiv & 3N_c^2+14N_c-9 \ .
\end{eqnarray}
In alternate choices for distinguishing ``$8_{1,2}$'' and
``$10_{1,2}$'', different factors appear.

\section{SU(3) Isoscalar Factors}

\begin{displaymath}
\mbox{\rm Isoscalar factors of the form }
\iso{``8\mbox{\rm ''}}{8}{\mu_\gamma}{I_1 Y_1}{I_2 Y_2}{I Y} :
\end{displaymath}

\renewcommand{\arraystretch}{1.75}

\begin{tabular}{|cccc|c|c}
\multicolumn{5}{c}{$Y=\frac{N_c}{3} \! + \! 1, \ I \! = \! 1$} \\
\hline
$I_1$, & $Y_1$; & $I_2$, & $Y_2$ & ``27'' & \ $\mu_\gamma$ \\
\hline
$\frac 1 2$, & $\frac{N_c}{3}$; & $\frac 1 2$, & $+1$ & $+1$ \\
\cline{1-5}
\end{tabular}
\hspace{2em}
\begin{tabular}{|cccc|c|c}
\multicolumn{5}{c}{$Y=\frac{N_c}{3} \! + \! 1$, $I \! = \! 0$} \\
\hline
$I_1$, & $Y_1$; & $I_2$, & $Y_2$ & ``$\overline{10}$'' & \
$\mu_\gamma$ \\
\hline
$\frac 1 2$, & $\frac{N_c}{3}$; & $\frac 1 2$, & $+1$ & $-1$ \\
\cline{1-5}
\end{tabular} \vspace{1em}

\begin{tabular}{|cccc|cc|c}
\multicolumn{6}{c}{$Y=\frac{N_c}{3}$, $I \! = \! \frac 3 2$} \\
\hline
$I_1$, & $Y_1$; & $I_2$, & $Y_2$ & ``27'' & ``10'' & \ $\mu_\gamma$ \\
\hline
$\frac 1 2$, & $\frac{N_c}{3}$; & 1, & 0 & $+\sqrt{\frac{2}{N_c+1}}$ &
$-\sqrt{\frac{N_c-1}{N_c+1}}$ \\
1, & $\frac{N_c}{3} \! - \! 1$; & $\frac 1 2$, & $+1$ &
$+\sqrt{\frac{N_c-1}{N_c+1}}$ & $+\sqrt{\frac{2}{N_c+1}}$ \\
\cline{1-6}
\end{tabular}
\hspace{2em}
\begin{tabular}
{|cccc|c|c}
\multicolumn{5}{c}{$Y=\frac{N_c}{3} \! - \! 1, \ I \! = \! 2$} \\
\hline
$I_1$, & $Y_1$; & $I_2$, & $Y_2$ & ``27'' & \ $\mu_\gamma$ \\
\hline
1, & $\frac{N_c}{3} \! - \! 1$; & 1 & 0 & $+\frac{2}{\sqrt{N_c+1}}$
\\
\cline{1-5}
\end{tabular} \vspace{1em}

\begin{tabular}{|cccc|cccc|c}
\multicolumn{8}{c}{$Y=\frac{N_c}{3}, \ I \! = \! \frac{1}{2}$} \\
\hline
$I_1$, & $Y_1$; & $I_2$, & $Y_2$ & ``27'' & ``$\overline{10}$'' &
``$8_1$'' & ``$8_2$'' & \ $\mu_\gamma$ \\
\hline
$\frac 1 2$, & $\frac{N_c}{3}$; & 1, & 0 &
$+\frac{1}{\sqrt{2(N_c+7)}}$ & $-\sqrt{\frac{3}{2(N_c+3)}}$ &
$+{\scriptstyle 3}\sqrt{\frac{(N_c-1)(N_c+3)}{2D}}$ &
$+\frac{N_c^2+8N_c+27}{\sqrt{2D(N_c+7)(N_c+3)}}$ \\
1, & $\frac{N_c}{3} \! - \! 1$; & $\frac 1 2$, & $+1$ &
$-\frac{1}{2}\sqrt{\frac{N_c-1}{N_c+7}}$ &
$-\frac{1}{2}\sqrt{\frac{3(N_c-1)}{N_c+3}}$ &
$-{\scriptstyle 3}\sqrt{\frac{N_c+3}{D}}$ &
$+\frac{2(2N_c+9)\sqrt{N_c-1}}{\sqrt{D(N_c+7)(N_c+3)}}$ \\
$\frac 1 2$, & $\frac{N_c}{3}$; & 0, & 0 &
$+\frac{3}{\sqrt{2(N_c+7)}}$ & $+\sqrt{\frac{3}{2(N_c+3)}}$ &
$-\sqrt{\frac{(N_c-1)(N_c+3)}{2D}}$ &
$+\frac{3(N_c^2+4N_c-1)}{\sqrt{2D(N_c+7)(N_c+3)}}$ \\
0, & $\frac{N_c}{3} \! - \! 1$; & $\frac 1 2$, & $+1$ & $+\frac 1 2
\sqrt{\frac{3(N_c+3)}{N_c+7}}$ & $-\frac 1 2$ &
$-\sqrt{\frac{3(N_c-1)}{D}}$ &
$-\frac{2(N_c+2)\sqrt{3}}{\sqrt{D(N_c+7)}}$ \\
\cline{1-8}
\end{tabular} \vspace{1em}

\begin{tabular}{|cccc|ccccc|c}
\multicolumn{9}{c}{$Y=\frac{N_c}{3} \! - \! 1, \ I \! = \! 1$} \\
\hline
$I_1$, & $Y_1$; & $I_2$, & $Y_2$ & ``27'' & ``$\overline{10}$'' &
``10'' & ``$8_1$'' & ``$8_2$'' & \ $\mu_\gamma$ \\
\hline
$\frac 1 2$, & $\frac{N_c}{3}$; & $\frac 1 2$, & $-1$ &
$+{\scriptstyle 2} \sqrt{\frac{2}{(N_c+7)(N_c+1)}}$ &
$+\frac{2}{\sqrt{(N_c+3)(N_c+1)}}$ & $-{\scriptstyle 2}
\sqrt{\frac{2(N_c-1)}{3(N_c+5)(N_c+1)}}$ &
$-\sqrt{\frac{6(N_c+3)}{D}}$ &
$+\frac{2(2N_c+9)\sqrt{2(N_c-1)}}{\sqrt{3D(N_c+7)(N_c+3)}}$ \\
$\frac 1 2$, & $\frac{N_c}{3} \! - \! 2$; & $\frac 1 2$, & $+1$ &
$+\sqrt{\frac{2(N_c+3)(N_c-1)}{3(N_c+7)(N_c+1)}}$ &
$-\sqrt{\frac{N_c-1}{3(N_c+1)}}$ & $+\frac 2 3
\sqrt{\frac{2(N_c+3)}{(N_c+5)(N_c+1)}}$ & $-{\scriptstyle 2N_c}
\sqrt{\frac{2}{D(N_c-1)}}$ & $-\frac{7N_c+9}{3}
\sqrt{\frac{2}{D(N_c+7)}}$ \\
1, & $\frac{N_c}{3} \! - \! 1$; & 1, & 0 & 0 &
$-\sqrt{\frac{2(N_c-1)}{(N_c+3)(N_c+1)}}$ &
$+\sqrt{\frac{N_c+1}{3(N_c+5)}}$ & $+{\scriptstyle (N_c-3)}
\sqrt{\frac{3(N_c+3)}{D(N_c-1)}}$ & $+{\scriptstyle (N_c+9)}
\sqrt{\frac{N_c+7}{3D(N_c+3)}}$ \\
1, & $\frac{N_c}{3} \! - \! 1$; & 0, & 0 &
$+\sqrt{\frac{6(N_c-1)}{(N_c+7)(N_c+1)}}$ &
$+\sqrt{\frac{3(N_c-1)}{(N_c+3)(N_c+1)}}$ & $+{\scriptstyle 2}
\sqrt{\frac{2}{(N_c+5)(N_c+1)}}$ & $-{\scriptstyle (N_c-7)}
\sqrt{\frac{N_c+3}{2D(N_c-1)}}$ & $+\frac{(N_c-3)(3N_c+13)}
{\sqrt{2D(N_c+7)(N_c+3)}}$ \\
0, & $\frac{N_c}{3} \! - \! 1$; & 1, & 0 &
$+\sqrt{\frac{2(N_c+3)}{(N_c+7)(N_c+1)}}$ & $-\frac{1}{\sqrt{N_c+1}}$
& $-\sqrt{\frac{2(N_c+3)(N_c-1)}{3(N_c+5)(N_c+1)}}$ & $+{\scriptstyle
(N_c+1)} \sqrt{\frac{3}{2D}}$ & $+{\scriptstyle (N_c-3)}
\sqrt{\frac{(N_c-1)}{6D(N_c+7)}}$ \\
\cline{1-9}
\end{tabular} \vspace{1em}

\begin{tabular}{|cccc|cccc|c}
\multicolumn{8}{c}{$Y=\frac{N_c}{3} \! - \! 1, \ I \! = \! 0$} \\
\hline
$I_1$, & $Y_1$; & $I_2$, & $Y_2$ & ``27'' & ``$8_1$'' & ``$8_2$'' &
``1'' & \ $\mu_\gamma$ \\
\hline
$\frac 1 2$, & $\frac{N_c}{3}$; & $\frac 1 2$, & $-1$ &
$+{\scriptstyle 2} \sqrt{\frac{3}{(N_c+7)(N_c+5)}}$ &
$+\sqrt{\frac{6(N_c-1)}{D}}$ &
$+\frac{2(N_c+2)\sqrt{6}}{\sqrt{D(N_c+7)}}$ &
$+\sqrt{\frac{N_c-1}{N_c+5}}$ \\
$\frac 1 2$, & $\frac{N_c}{3} \! - \! 2$; & $\frac 1 2$, & $+1$ &
$-\sqrt{\frac{(N_c+3)(N_c-1)}{(N_c+7)(N_c+5)}}$ & $-{\scriptstyle 6}
\sqrt{\frac{2}{D(N_c+3)}}$ & $+{\scriptstyle 5}
\sqrt{\frac{2(N_c+3)(N_c-1)}{D(N_c+7)}}$ & $-{\scriptstyle 2}
\sqrt{\frac{3}{(N_c+5)(N_c+3)}}$ \\
1, & $\frac{N_c}{3} \! - \! 1$; & 1, & 0 &
$-\sqrt{\frac{N_c-1}{(N_c+7)(N_c+5)}}$ & $-\frac{3(N_c+1)}{\sqrt{2D}}$
& $-{\scriptstyle (N_c-3)} \sqrt{\frac{N_c-1}{2D(N_c+7)}}$ &
$+\sqrt{\frac{3}{N_c+5}}$ \\
0, & $\frac{N_c}{3} \! - \! 1$; & 0, & 0 &
$+{\scriptstyle 3} \sqrt{\frac{N_c+3}{(N_c+7)(N_c+5)}}$ &
$-{\scriptstyle (N_c+9)} \sqrt{\frac{N_c-1}{2D(N_c+3)}}$ &
$+{\scriptstyle 3(N_c-3)} \sqrt{\frac{N_c+3}{2D(N_c+7)}}$ &
$-\sqrt{\frac{3(N_c-1)}{(N_c+5)(N_c+3)}}$ \\
\cline{1-8}
\end{tabular} \vspace{1em}

\begin{tabular}{|cccc|cc|c}
\multicolumn{6}{c}{$Y=\frac{N_c}{3} \! - \! 2, \ I \! = \! \frac 3 2$}
\\ \hline
$I_1$, & $Y_1$; & $I_2$, & $Y_2$ & ``27'' & ``$\overline{10}$'' & \
$\mu_\gamma$ \\
\hline
$\frac 1 2$, & $\frac{N_c}{3} \! - \! 2$; & 1, & 0 &
$+\sqrt{\frac{10(N_c+3)}{3(N_c+7)(N_c+1)}}$ &
$-\sqrt{\frac{2}{N_c+1}}$ \\
1, & $\frac{N_c}{3} \! - \! 1$; & $\frac 1 2$, & $-1$ &
$+{\scriptstyle 2} \sqrt{\frac{5}{(N_c+7)(N_c+1)}}$ &
$+{\scriptstyle 2} \sqrt{\frac{3}{(N_c+3)(N_c+1)}}$ \\
\cline{1-6}
\end{tabular} \vspace{1em}

\begin{tabular}{|cccc|cccc|c}
\multicolumn{8}{c}{$Y=\frac{N_c}{3} \! - \! 2, \ I \! = \! \frac 1 2$}
\\ \hline
$I_1$, & $Y_1$; & $I_2$, & $Y_2$ & ``27'' & ``10'' & ``$8_1$'' &
``$8_2$'' & \ $\mu_\gamma$ \\
\hline
$\frac 1 2$, & $\frac{N_c}{3} \! - \! 2$; & 1, & 0 & $-{\scriptstyle
2} \sqrt{\frac{(N_c+3)(N_c-1)}{3(N_c+7)(N_c+5)(N_c+1)}}$ &
$+\frac{2N_c \sqrt{2}}{3\sqrt{(N_c+5)(N_c+1)}}$ &
$-\frac{(N_c^2+4N_c+15)}{\sqrt{2D(N_c+3)(N_c-1)}}$ &
$-\frac{N_c-33}{3} \sqrt{\frac{N_c+3}{2D(N_c+7)}}$ \\
1, & $\frac{N_c}{3} \! - \! 1$; & $\frac 1 2$, & $-1$ &
$+{\scriptstyle 2} \sqrt{\frac{2(N_c-1)}{(N_c+7)(N_c+5)(N_c+1)}}$ &
$+{\scriptstyle 2} \sqrt{\frac{N_c+3}{3(N_c+5)(N_c+1)}}$ &
$+\frac{2N_c \sqrt{3}}{\sqrt{D(N_c-1)}}$ &
$+\frac{7N_c+9}{\sqrt{3D(N_c+7)}}$ \\
$\frac 1 2$, & $\frac{N_c}{3} \! - \! 2$; & 0, & 0 & $+{\scriptstyle
2} \sqrt{\frac{3(N_c+3)(N_c-1)}{(N_c+7)(N_c+5)(N_c+1)}}$ &
$+{\scriptstyle 2} \sqrt{\frac{2}{(N_c+5)(N_c+1)}}$ &
$-\frac{(N_c^2+8N_c-21)}{\sqrt{2D(N_c+3)(N_c-1)}}$ & $-{\scriptstyle
(3N_c-19)} \sqrt{\frac{N_c+3}{2D(N_c+7)}}$ \\
0, & $\frac{N_c}{3} \! - \! 1$; & $\frac 1 2$, & $-1$ &
$+{\scriptstyle 2} \sqrt{\frac{6(N_c+3)}{(N_c+7)(N_c+5)(N_c+1)}}$ &
$-{\scriptstyle 2} \sqrt{\frac{N_c-1}{(N_c+5)(N_c+1)}}$ &
$-\frac{6}{\sqrt{D(N_c+3)}}$ & $+{\scriptstyle 5}
\sqrt{\frac{(N_c+3)(N_c-1)}{D(N_c+7)}}$ \\
\cline{1-8}
\end{tabular} \vspace{1em}

\begin{tabular}{|cccc|c|c}
\multicolumn{5}{c}{$Y=\frac{N_c}{3} \! - \! 3, \ I \! = \! 1$} \\
\hline
$I_1$, & $Y_1$; & $I_2$, & $Y_2$ & ``27'' & \ $\mu_\gamma$ \\
\hline
$\frac 1 2$, & $\frac{N_c}{3} \! - \! 2$; & $\frac 1 2$, & $-1$ &
$+{\scriptstyle 4} \sqrt{\frac{10(N_c+3)}{3(N_c+7)(N_c+5)(N_c+1)}}$ \\
\cline{1-5}
\end{tabular}
\hspace{2em}
\begin{tabular}{|cccc|c|c}
\multicolumn{5}{c}{$Y=\frac{N_c}{3} \! - \! 3, \ I \! = \! 0$} \\
\hline
$I_1$, & $Y_1$; & $I_2$, & $Y_2$ & ``10'' & \ $\mu_\gamma$ \\
\hline
$\frac 1 2$, & $\frac{N_c}{3} \! - \! 2$; & $\frac 1 2$, & $-1$ &
$+{\scriptstyle 2} \sqrt{\frac{2}{N_c+5}}$ \\
\cline{1-5}
\end{tabular} 
\newpage

\renewcommand{\arraystretch}{1.00}
\begin{displaymath}
\mbox{\rm Isoscalar factors of the form }
\iso{``10\mbox{\rm ''}}{8}{\mu_\gamma}{I_1 Y_1}{I_2 Y_2}{I Y} :
\end{displaymath}
\renewcommand{\arraystretch}{1.75}

\begin{tabular}{|cccc|c|c}
\multicolumn{5}{c}{$Y=\frac{N_c}{3} \! + \! 1, \ I \! = \! 2$} \\
\hline
$I_1$, & $Y_1$; & $I_2$, & $Y_2$ & ``35'' & \ $\mu_\gamma$ \\
\hline
$\frac 3 2$, & $\frac{N_c}{3}$; & $\frac 1 2$, & $+1$ & $+1$ \\
\cline{1-5}
\end{tabular}
\hspace{2em}
\begin{tabular}{|cccc|c|c}
\multicolumn{5}{c}{$Y=\frac{N_c}{3} \! + \! 1$, $I \! = \! 1$} \\
\hline
$I_1$, & $Y_1$; & $I_2$, & $Y_2$ & ``27'' & \ $\mu_\gamma$ \\
\hline
$\frac 3 2$, & $\frac{N_c}{3}$; & $\frac 1 2$, & $+1$ & $-1$ \\
\cline{1-5}
\end{tabular}
\hspace{2em}
\begin{tabular}{|cccc|c|c}
\multicolumn{5}{c}{$Y=\frac{N_c}{3}$, $I \! = \! \frac 5 2$} \\
\hline
$I_1$, & $Y_1$; & $I_2$, & $Y_2$ & ``35'' & \ $\mu_\gamma$ \\
\hline
$\frac 3 2$, & $\frac{N_c}{3}$; & 1, & 0 &
$+\sqrt{\frac{2}{N_c-1}}$ \\
\cline{1-5}
\end{tabular} \vspace{1em}

\begin{tabular}{|cccc|ccc|c}
\multicolumn{7}{c}{$Y=\frac{N_c}{3}, \ I \! = \! \frac{3}{2}$} \\
\hline
$I_1$, & $Y_1$; & $I_2$, & $Y_2$ & ``35'' & ``27'' & ``$10_2$'' & \
$\mu_\gamma$ \\
\hline
$\frac 3 2$, & $\frac{N_c}{3}$; & 1, & 0 & $+\frac 1 2
\sqrt{\frac{3}{N_c+9}}$ & $-\frac 1 2 \sqrt{\frac{5}{N_c+1}}$ &
$+\frac{N_c^2+8N_c+27}{\sqrt{2\tilde{D}(N_c+9)(N_c+1)}}$ \\
$\frac 3 2$, & $\frac{N_c}{3}$; & 0, & 0 &
$+\frac{1}{2}\sqrt{\frac{15}{N_c+9}}$ & $+\frac{3}{2\sqrt{N_c+1}}$ &
$+\frac{(N_c^2+4N_c-9)\sqrt{5}}{\sqrt{2\tilde{D}(N_c+9)(N_c+1)}}$ \\
1, & $\frac{N_c}{3} \! - \! 1$; & $\frac 1 2$, & $+1$ &
$+\frac{1}{4}\sqrt{\frac{15(N_c+5)}{N_c+9}}$ & $-\frac{1}{4}
\sqrt{\frac{N_c+5}{N_c+1}}$ & $-{\scriptstyle N_c}
\sqrt{\frac{10(N_c+5)}{\tilde{D}(N_c+9)(N_c+1)}}$
\\
\cline{1-7}
\end{tabular} \vspace{1em}

\begin{tabular}{|cccc|cc|c}
\multicolumn{6}{c}{$Y=\frac{N_c}{3}$, $I \! = \! \frac 1 2$} \\
\hline
$I_1$, & $Y_1$; & $I_2$, & $Y_2$ & ``27'' & ``8'' & \ $\mu_\gamma$ \\
\hline
$\frac 3 2$, & $\frac{N_c}{3}$; & 1, & 0 & $-\sqrt{\frac{2}{N_c+7}}$ &
$-\sqrt{\frac{N_c+5}{N_c+7}}$ \\
1, & $\frac{N_c}{3} \! - \! 1$; & $\frac 1 2$, & $+1$ &
$-\sqrt{\frac{N_c+5}{N_c+7}}$ & $+\sqrt{\frac{2}{N_c+7}}$ \\
\cline{1-6}
\end{tabular}
\hspace{2em}
\begin{tabular}
{|cccc|cc|c}
\multicolumn{6}{c}{$Y=\frac{N_c}{3} \! - \! 1, \ I \! = \! 2$} \\
\hline
$I_1$, & $Y_1$; & $I_2$, & $Y_2$ & ``35'' & ``27'' & \
$\mu_\gamma$ \\
\hline
1, & $\frac{N_c}{3} \! - \! 1$; & 1 & 0 & $+\frac{3}{2}
\sqrt{\frac{N_c+5}{(N_c+9)(N_c-1)}}$ & $-\frac{1}{2}
\sqrt{\frac{N_c+5}{(N_c+1)(N_c-1)}}$ \\
$\frac 3 2$, & $\frac{N_c}{3}$; & $\frac 1 2$, & $-1$ &
$+\sqrt{\frac{6}{(N_c+9)(N_c-1)}}$ &
$+\sqrt{\frac{6}{(N_c+1)(N_c-1)}}$ \\
\cline{1-6}
\end{tabular} \vspace{1em}

\begin{tabular}{|cccc|cccc|c}
\multicolumn{8}{c}{$Y=\frac{N_c}{3} \! - \! 1, \ I \! = \! 1$} \\
\hline
$I_1$, & $Y_1$; & $I_2$, & $Y_2$ & ``35'' & ``27'' & ``$10_2$'' &
``8'' & \ $\mu_\gamma$ \\
\hline
1, & $\frac{N_c}{3} \! - \! 1$; & 1, & 0 & $+\frac 1 2
\sqrt{\frac{5(N_c+5)}{(N_c+9)(N_c+7)}}$ & $-\frac 3 2
\sqrt{\frac{N_c+5}{(N_c+7)(N_c+1)}}$
& $+\frac{(N_c^2+4N_c+27)\sqrt{5}}{2\sqrt{3\tilde{D}(N_c+9)(N_c+1)}}$
& $-\frac{N_c+1}{\sqrt{6(N_c+7)(N_c-1)}}$ \\
1, & $\frac{N_c}{3} \! - \! 1$; & 0,  & 0 &
$+\sqrt{\frac{15(N_c+5)}{2(N_c+9)(N_c+7)}}$ &
$+\sqrt{\frac{3(N_c+5)}{2(N_c+7)(N_c+1)}}$ &
$+\frac{(N_c+3)(N_c-3)\sqrt{5}}{\sqrt{2\tilde{D}(N_c+9)(N_c+1)}}$ &
$-\frac{2}{\sqrt{(N_c+7)(N_c-1)}}$ \\
$\frac 1 2$, & $\frac{N_c}{3} \! - \! 2$; & $\frac 1 2$, & $+1$ &
$+\sqrt{\frac{5(N_c+5)(N_c+3)}{6(N_c+9)(N_c+7)}}$ &
$-\sqrt{\frac{(N_c+5)(N_c+3)}{6(N_c+7)(N_c+1)}}$ &
$-\frac{4N_c}{3}\sqrt{\frac{10(N_c+3)}{\tilde{D}(N_c+9)(N_c+1)}}$ &
$+\frac 2 3 \sqrt{\frac{N_c+3}{(N_c+7)(N_c-1)}}$ \\
$\frac 3 2$, & $\frac{N_c}{3}$; & $\frac 1 2$, & $-1$ &
$+\sqrt{\frac{10}{(N_c+9)(N_c+7)}}$, &
$+\sqrt{\frac{2}{(N_c+7)(N_c+1)}}$; & $+{\scriptstyle 2N_c}
\sqrt{\frac{10(N_c+5)}{3\tilde{D}(N_c+9)(N_c+1)}}$ & $+{\scriptstyle
2} \sqrt{\frac{N_c+5}{3(N_c+7)(N_c-1)}}$ \\
\cline{1-8}
\end{tabular} \vspace{1em}

\begin{tabular}{|cccc|cc|c}
\multicolumn{6}{c}{$Y=\frac{N_c}{3} \! - \! 1$, $I \! = \! 0$} \\
\hline
$I_1$, & $Y_1$; & $I_2$, & $Y_2$ & ``27'' & ``8'' & \ $\mu_\gamma$ \\
\hline
1, & $\frac{N_c}{3} \! - \! 1$; & 1, & 0 & $-\frac{2}{\sqrt{N_c+7}}$ &
$-\sqrt{\frac{N_c+3}{N_c+7}}$ \\
$\frac 1 2$, & $\frac{N_c}{3} \! - \! 2$; & $\frac 1 2$, & $+1$ &
$-\sqrt{\frac{N_c+3}{N_c+7}}$ & $+\frac{2}{\sqrt{N_c+7}}$ \\
\cline{1-6}
\end{tabular} \vspace{1em}

\begin{tabular}
{|cccc|cc|c}
\multicolumn{6}{c}{$Y=\frac{N_c}{3} \! - \! 2, \ I \! = \! \frac 3 2$} \\
\hline
$I_1$, & $Y_1$; & $I_2$, & $Y_2$ & ``35'' & ``27'' & \ $\mu_\gamma$ \\
\hline
$\frac 1 2$, & $\frac{N_c}{3} \! - \! 2$; & 1 & 0 &
$+\sqrt{\frac{5(N_c+5)(N_c+3)}{2(N_c+9)(N_c+7)(N_c-1)}}$ &
$-\sqrt{\frac{5(N_c+5)(N_c+3)}{6(N_c+7)(N_c+1)(N_c-1)}}$ \\
1, & $\frac{N_c}{3} \! - \! 1$; & $\frac 1 2$, & $-1$ &
$+\sqrt{\frac{15(N_c+5)}{(N_c+9)(N_c+7)(N_c-1)}}$ &
$+\sqrt{\frac{5(N_c+5)}{(N_c+7)(N_c+1)(N_c-1)}}$ \\
\cline{1-6}
\end{tabular} \vspace{1em}

\begin{tabular}{|cccc|cccc|c}
\multicolumn{8}{c}{$Y=\frac{N_c}{3} \! - \! 2, \ I \! = \! \frac 1 2$}
\\
\hline
$I_1$, & $Y_1$; & $I_2$, & $Y_2$ & ``35'' & ``27'' & ``$10_2$'' &
``8'' & \ $\mu_\gamma$ \\
\hline
$\frac 1 2$, & $\frac{N_c}{3} \! - \! 2$; & 1, & 0 & $+\frac 1 2
\sqrt{\frac{5(N_c+3)}{(N_c+9)(N_c+7)}}$ &
$-\frac 7 2 \sqrt{\frac{N_c+3}{3(N_c+7)(N_c+1)}}$ &
$+\frac{(N_c^2+27) \sqrt{5}}{3\sqrt{2\tilde{D} (N_c+9)(N_c+1)}}$ &
$-\frac{2N_c}{3\sqrt{(N_c+7)(N_c+1)}}$ \\
$\frac 1 2$, & $\frac{N_c}{3} \! - \! 2$; & 0, & 0 &
$+\frac{3}{2}\sqrt{\frac{5(N_c+3)}{(N_c+9)(N_c+7)}}$ &
$+\frac{1}{2}\sqrt{\frac{3(N_c+3)}{(N_c+7)(N_c+1)}}$ &
$+\frac{(N_c^2-4N_c-9)\sqrt{5}}{\sqrt{2\tilde{D}(N_c+9)(N_c+1)}}$ &
$-\frac{2}{\sqrt{(N_c+7)(N_c-1)}}$ \\
0, & $\frac{N_c}{3} \! - \! 3$; & $\frac 1 2$, & $+1$ & $+\frac{1}{2}
\sqrt{\frac{5(N_c+3)(N_c+1)}{2(N_c+9)(N_c+7)}}$ &
$-\frac 1 2 \sqrt{\frac{3(N_c+3)}{2(N_c+7)}}$ &
$-{\scriptstyle 2N_c} \sqrt{\frac{5}{\tilde{D}(N_c+9)}}$ &
$+\sqrt{\frac{2(N_c+1)}{(N_c+7)(N_c-1)}}$ \\
1, & $\frac{N_c}{3} \! - \! 1$; & $\frac 1 2$, & $-1$ &
$+\sqrt{\frac{30}{(N_c+9)(N_c+7)}}$ &
$+\sqrt{\frac{2}{(N_c+7)(N_c+1)}}$ & $+{\scriptstyle 4N_c}
\sqrt{\frac{5(N_c+3)}{3\tilde{D}(N_c+9)(N_c+1)}}$ &
$+\sqrt{\frac{2(N_c+3)}{3(N_c+7)(N_c-1)}}$ \\
\cline{1-8}
\end{tabular} \vspace{1em}

\begin{tabular}{|cccc|cc|c}
\multicolumn{6}{c}{$Y=\frac{N_c}{3} \! - \! 3$, $I \! = \! 1$} \\
\hline
$I_1$, & $Y_1$; & $I_2$, & $Y_2$ & ``35'' & ``27'' & \ $\mu_\gamma$ \\
\hline
0, & $\frac{N_c}{3} \! - \! 3$; & 1, & 0 &
$+\sqrt{\frac{5(N_c+3)(N_c+1)}{2(N_c+9)(N_c+7)(N_c-1)}}$ &
$-\sqrt{\frac{5(N_c+3)}{2(N_c+7)(N_c-1)}}$ \\
$\frac 1 2$, & $\frac{N_c}{3} \! - \! 2$; & $\frac 1 2$, & $-1$ &
$+\sqrt{\frac{30(N_c+3)}{(N_c+9)(N_c+7)(N_c-1)}}$ &
$+\sqrt{\frac{10(N_c+3)}{3(N_c+7)(N_c+1)(N_c-1)}}$ \\
\cline{1-6}
\end{tabular} \vspace{1em}

\begin{tabular}
{|cccc|cc|c}
\multicolumn{6}{c}{$Y=\frac{N_c}{3} \! - \! 3, \ I \! = \! 0$} \\
\hline
$I_1$, & $Y_1$; & $I_2$, & $Y_2$ & ``35'' & ``$10_2$'' & \
$\mu_\gamma$ \\
\hline
0, & $\frac{N_c}{3} \! - \! 3$; & 0 & 0 &
$+\sqrt{\frac{15(N_c+1)}{(N_c+9)(N_c+7)}}$ & $+{\scriptstyle (N_c-9)}
\sqrt{\frac{5(N_c+1)}{2\tilde{D}(N_c+9)}}$ \\
$\frac 1 2$, & $\frac{N_c}{3} \! - \! 2$; & $\frac 1 2$, & $-1$ &
$+{\scriptstyle 2} \sqrt{\frac{15}{(N_c+9)(N_c+7)}}$ &
$+{\scriptstyle 2N_c} \sqrt{\frac{10}{\tilde{D}(N_c+9)}}$ \\
\cline{1-6}
\end{tabular}
\hspace{1.5em}
\begin{tabular}{|cccc|c|c}
\multicolumn{5}{c}{$Y=\frac{N_c}{3} \! - \! 4, \ I \! = \! \frac 1 2$}
\\
\hline
$I_1$, & $Y_1$; & $I_2$, & $Y_2$ & ``35'' & \ $\mu_\gamma$ \\
\hline
0, & $\frac{N_c}{3} \! - \! 3$; & $\frac 1 2$, & $-1$ &
$+{\scriptstyle 2} \sqrt{\frac{15(N_c+1)}{(N_c+9)(N_c+7)(N_c-1)}}$ \\
\cline{1-5}
\end{tabular} 

\end{document}